\documentclass[prl,preprint,superscriptaddress,longbibliography]{revtex4-1}
\AtBeginDocument{%
    \newwrite\bibnotes
    \def\bibnotesext{Notes.bib}
    \immediate\openout\bibnotes=\jobname\bibnotesext
    \immediate\write\bibnotes{@CONTROL{REVTEX41Control}}
    \immediate\write\bibnotes{@CONTROL{%
    apsrev41Control,author="08",editor="1",pages="1",title="0",year="1"}}
     \if@filesw
     \immediate\write\@auxout{\string\citation{apsrev41Control}}%
    \fi
}%

\usepackage{graphicx}
\usepackage{hyperref}
\usepackage{amsmath}
\usepackage[usenames,dvipsnames]{color}
\usepackage{amssymb}
\usepackage{xcolor}
\usepackage[colorinlistoftodos]{todonotes}

\begin{document}

\title{Engineered electronic states in atomically precise artificial lattices and graphene nanoribbons}

\author{Linghao Yan}
\affiliation{Department of Applied Physics, Aalto University School of Science, P.O. Box 15100, 00076 Aalto, Finland}

\author{Peter Liljeroth}
\email{Email: peter.liljeroth@aalto.fi}
\affiliation{Department of Applied Physics, Aalto University School of Science, P.O. Box 15100, 00076 Aalto, Finland}

\date{\today}
\begin{abstract}
The fabrication of atomically precise structures with designer electronic properties is one of the emerging topics in condensed matter physics. The required level of structural control can either be reached through atomic manipulation using the tip of a scanning tunneling microscope (STM) or by bottom-up chemical synthesis. In this review, we focus on recent progress in constructing novel, atomically precise artificial materials: artificial lattices built through atom manipulation and graphene nanoribbons (GNRs) realized by on-surface synthesis. We summarize the required theoretical background and the latest experiments on artificial lattices, topological states in one-dimensional lattices, experiments on graphene nanoribbons and graphene nanoribbon heterostructures, and topological states in graphene nanoribbons. Finally, we conclude our review with an outlook to designer quantum materials with engineered electronic structure.
\end{abstract}
\maketitle

\section{Introduction}

Creating and studying nanostructures where each atom is in a well-defined, pre-determined position is currently being vigorously pursued within condensed-matter physics and materials chemistry research communities. This type of atomic-level control cannot be achieved by standard high-resolution lithographic techniques such as electron beam lithography. At the same time, it would obviously open a new path to constructing materials with engineered electronic properties and there are approaches available that are making it a reality. The required precision can either be reached by positioning every atom of the nanostructure by the tip of a scanning probe microscope, or by clever bottom-up chemical strategies that allow direct synthesis of well-defined structures from molecular precursors. Atom manipulation by scanning tunnelling microscope (STM) or atomic force microscope (AFM) can be used on single-crystal surfaces to place individual atoms on the desired positions to construct proof-of-principle -type individual nanostructures \cite{Eigler1990atom_manipulation,Eigler1991AnMicroscope,Stroscio1991AtomicMicroscope,Ternes2008TheSurface}. At the same time, the electronic properties of the structures can be characterized in-situly by STM by measuring the d$I$/d$V$ signal (tunneling conductance) that is directly proportional to the local density of states at the position of the STM tip \cite{Chen2007book,Ervasti2017reviewstm}. This has been used to demonstrate, for example, quantum confinement of surface state electrons in so-called quantum corrals \cite{Crommie1993ImagingGas,Manoharan2000QuantumStructure,Moon2009QuantumGas}, computation using molecular cascades \cite{Heinrich2002MoleculeCascades}, and the formation of artificial graphene by confining a surface state electron gas into a honeycomb lattice \cite{Gomes2012DesignerGraphene}. In addition to scanning probe -based atomic manipulation, there are significant efforts to use focused electron beams to sculpt materials with atomic precision \cite{Jesse2016,Dyck2017,Susi2017,Susi2017aa,Zhao2018aa,Hudak2018aa,Tripathi2018aa,Su2019aa,Mustonen2019aa,Dyck2019}.

This scanning probe -based approach naturally cannot be used to mass-produce nanostructures but is limited to prototyping the targeted physical effects and mechanisms. A way forward towards larger scale production of materials with desired, engineered properties is to use suitable bottom-up approaches. These are part of the growing field of on-surface synthesis, which includes, for example, the formation of metal-organic networks with controlled lattice structures \cite{Barth2007MolecularSurfaces,Lobo-Checa2009BandSurface,Dong2016Self-assemblySurfaces} and chemical synthesis of atomically well-defined graphene nanoribbons (GNRs) based on molecular precursors \cite{Cai2010AtomicallyNanoribbons,Ruffieux2016On-surfaceTopology,Talirz2016On-SurfaceNanoribbons}. The beauty of these approaches is that the geometry and the symmetry of the structures are uniquely determined by the molecular precursor allowing for an unprecedented level of control and tunability \cite{Cai2010AtomicallyNanoribbons,vanderLit2013SuppressionAtom,Chen2013_ACSNano,Ruffieux2016On-surfaceTopology,Cai2014GrapheneHeterojunctions,Chen2015MolecularHeterojunctions,Kimouche2015Ultra-narrowNanoribbons,Talirz2016On-SurfaceNanoribbons}. In the case of the GNRs, this includes being able to predetermine the atomic width of the ribbon and its precise edge structure.

We will review both approaches to achieve novel artificial materials with engineered electronic properties. We will start by discussing the relevant key concepts required for understanding the cutting-edge experimental results that have been reached in recent years. For example, the control of the lattice structure gives complete control over the resulting band structure; \emph{e.g.}, creating a lattice with the honeycomb symmetry will mimic the extraordinary electronic properties of graphene that arise from the symmetry of the hexagonal lattice with a two-atom unit cell \cite{Gomes2012DesignerGraphene,CastroNeto2009TheGraphene}. In addition to band structure engineering, recent work has suggested ways to build topological properties into artificial systems and we will give an overview of the simple models that can capture such effects. These effects can also be realized in chemically synthesized graphene nanoribbons and armed with the understanding of the underlying physical mechanisms, we will review the most exciting experimental demonstrations of lattice engineering and topological states in artificial lattices and the results on graphene nanoribbons and nanoribbon heterostructures with predetermined geometry and properties. 

\section{Artificial lattices}

\textbf{Tight-binding model for two-dimensional lattices.} In considering the electronic structure of artificial lattices, we need some theoretical, conceptual understanding of how the lattice symmetry is connected to the electronic structure. This can be offered by the tight-binding (TB) theory, which considers the band structure arising from electrons hopping between localized atomic orbitals $| \Psi _ { i }\rangle$ \cite{Hoffmann1989book,Thomas2019perspective}. This can be expressed by the Hamiltonian consisting of a hopping term $t_{ij}$ between sites $i$ and $j$
\begin{eqnarray}
H = \sum _ { i } \epsilon _ { i } c _ { i } ^ { \dagger } c _ { i } - \sum _ { \langle i , j \rangle  } t_{ij} \left( c _ { i } ^ { \dagger } c _ { j } + H . c . \right)
\end{eqnarray}
where $\epsilon _ { i }$ is the on-site energy and $c^\dagger_{i} , c_{j}$ are the creation and annihilation operators. In principle, the summation runs over all atom pairs $\langle i , j \rangle $, but it is often sufficient to restrict this to the nearest- and next-nearest neighbours. The eigenenergies and eigenstates of the system can then be calculated by diagonalizing the Hamiltonian.
\begin{figure}[!b]
    \centering
    \includegraphics[width=.95\textwidth]{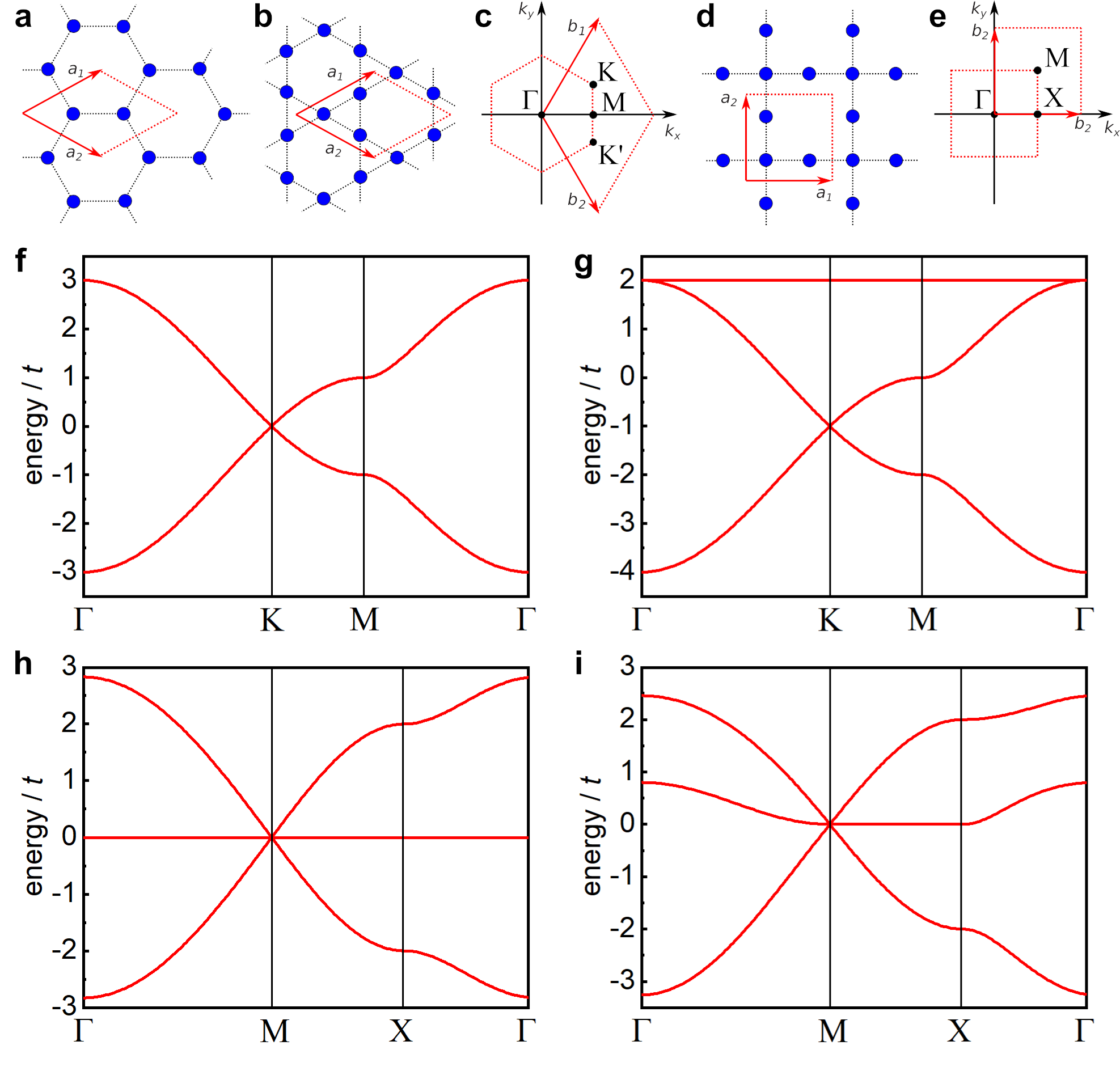}
    \caption{(a-e) Lattice structures and the Brillouin zones of the honeycomb (a), kagome (b) and Lieb (d) lattices. (f-h) Corresponding band structures calculated with the nearest-neighbour tight-binding model for the honeycomb (f), kagome (g) and Lieb (h) lattices. (i) The Lieb lattice band structure when second nearest-neighbour hoppings are introduced ($t'=0.2t$). The energies have been scaled by the nearest neighbour hopping strength $t$.}
    \label{fig:lattices}
\end{figure}

The power of the TB model can be introduced by considering a couple of lattices shown in Fig.~\ref{fig:lattices}a-e as examples. The resulting band structures are shown in Fig.~\ref{fig:lattices}f-i. It is well-known that the exciting electronic properties of graphene are intimately linked to its honeycomb lattice with a two-atom unit cell (Fig.~\ref{fig:lattices}a) \cite{CastroNeto2009TheGraphene}. The band structure can be very well approximated by the nearest-neighbour (NN) TB model; while quantitative agreement with \emph{e.g.}~density-functional theory (DFT) calculations can be reached by considering hoppings up to the third-nearest neighbours, the formation of the Dirac cones in the band structure and the linear dispersion around the $K$ points (at the corners of the Brillouin zone) is naturally already present in the NN model (Fig.~\ref{fig:lattices}f). This is a generic property of any honeycomb lattice and makes it possible to create ``artificial graphene'', that is, creating engineered systems, where electrons are confined onto a honeycomb lattice. This game can obviously be taken further and there are other lattice geometries that have the potential for hosting exotic electronic phases, but are not readily found in nature. For example, the kagome and Lieb lattices (see Fig.~\ref{fig:lattices}b and d) have the same Dirac band structure as the honeycomb lattice, but with an additional flat band pinned to the top (or bottom) of the Dirac band (kagome lattice, Fig.~\ref{fig:lattices}g) or located directly at the Dirac point (Lieb lattice,  Fig.~\ref{fig:lattices}h). The flat bands are interesting as they are prone to electronic instabilities and spontaneous symmetry breaking at (near) half filling. Depending on whether the electron-electron interactions are attractive or repulsive, this would result in the system becoming superconducting or magnetic, respectively \cite{Mielke1999_PRL,lieb1989theorems,Peotta2015SuperfluidityBands,lothman2017_PRB,Leykam2018ArtificialExperiments}. Finally, the presence of a flat band depends not only on the lattice symmetry but also on which hoppings are included in the model. For example, in the case of a Lieb lattice, the flat band becomes dispersive if the next-nearest neighbour hoppings are included (Fig.~\ref{fig:lattices}i).

\begin{figure}[!b]
    \centering
	\includegraphics[width=.95\textwidth]{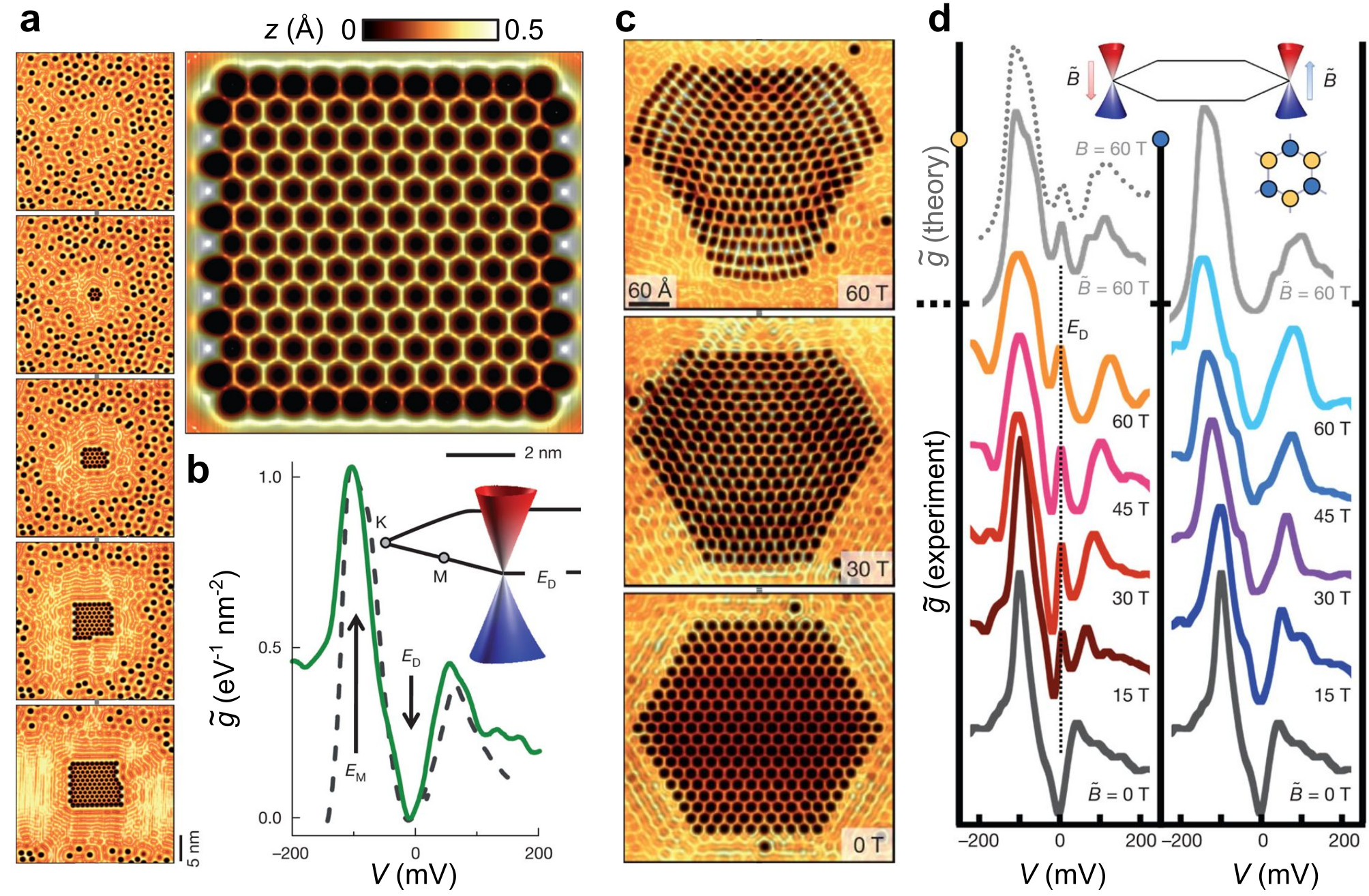}
	\caption{Experiments on artificial graphene. (a) Cu(111) surface state electrons are patterned by CO molecules ordered through STM lateral manipulation. (b) d$I$/d$V$ spectroscopy reveals the appearance of a Dirac cone in the band structure in the patterned area. (c) The effect of strain can be mimicked by continuously modulating the lattice spacing of the artificial graphene. The effective pseudomagnetic field strengths are indicated in the panels. (d) d$I$/d$V$ spectroscopy shows the formation of Landau levels due to the pseudomagnetic field generated by strain. Adapted  by permission from Springer Nature: Ref.~\citenum{Gomes2012DesignerGraphene}, Copyright (2012).}
	\label{fig:art_graphene}
\end{figure}

\textbf{Experiments on artificial lattices.} The idealized concepts presented in the previous section on band structure engineering can be realized in artificial lattices by STM manipulation \cite{Eigler1991AnMicroscope,Stroscio1991AtomicMicroscope,Hla2014Atom-by-atomAssembly}. This has been achieved by ``patterning'' the 2D electron gas formed by the surface state electrons on \emph{e.g.}~Cu(111) surface by adsorbates. The surface state electrons scatter from adsorbates which allows confining them into various structures \cite{Crommie1993ImagingGas,Hasegawa1993DirectSpectroscopy,Burgi1998ConfinementResonators,Negulyaev2008DirectDiffusion}. Previous work has shown that artificial quantum corrals, quantum mirages and quantum holographic encoding can be constructed through the confinement of electrons \cite{Crommie1993ConfinementSurface,Heller1994ScatteringCorrals,Manoharan2000QuantumStructure,Moon2009QuantumGas}. 

The exciting electronic properties of massless Dirac fermions in graphene have motivated research into artificial honeycomb lattices (artificial graphene) by different means such as ultra-cold atomic gases, phononic and photonic lattices, and two-dimensional electron gases in semiconductor nanopatterns \cite{Polini2013ArtificialPhotons,Yu2016SurfaceGraphene,Park2009bb}.
In particular, Gomes et al.~\cite{Gomes2012DesignerGraphene} realized artificial graphene in a condensed-matter system by manipulating carbon monoxide (CO) molecules  over the 2D electron gas on Cu(111) surface (Fig.~\ref{fig:art_graphene}a). When the CO molecules are patterned into a triangular repulsive potential array, the 2D electron gas is confined onto a honeycomb lattice and will mimic the behaviour of massless Dirac fermions (Fig.~\ref{fig:art_graphene}b). Due to the advantage of the atomically precise manipulation by STM, various kinds of artificial graphene structures were fabricated and investigated in this work: As the doping level can be modified by tuning the lattice parameters which will change the electron count per superlattice unit cell, a p-n-p junction is easily performed by combing lattices with smaller and larger spacing. Both pseudospin-breaking and pseudospin-conserving defects were fabricated. While C-site defect breaking the sublattice symmetry does not conserve the pseudospin, an empty-site vacancy preserves the sublattice symmetry and pseudospin conservation. Artificial graphene under Kekul\'e distortion breaks the bond symmetry and therefore opens a gap at the Dirac point. Finally, Gomes et al.~demonstrated the effect of triaxial strain in artificial graphene. In real graphene this would generate a pseudomagnetic field resulting in Landau levels and quantum Hall phases that have been observed in graphene nanobubbles \cite{Lee2008MeasurementGraphene}. This effect can be simulated in artificial graphene by moving the scatterers such that the couplings between the sites are modulated to mimic the effect of strain on the bond lengths in real graphene (Fig.~\ref{fig:art_graphene}c and d). Finally, the realization of artificial graphene is not limited to the use of CO molecules and further experiments have demonstrated using arrays of coronene molecules to successfully build artificial armchair and zigzag graphene nanoribbons and graphene with 558 line defects and Stone-Wales defects \cite{Wang2014ManipulationGas}. 
\begin{figure}[!b]
	\centering
	\includegraphics[width=0.95\textwidth]{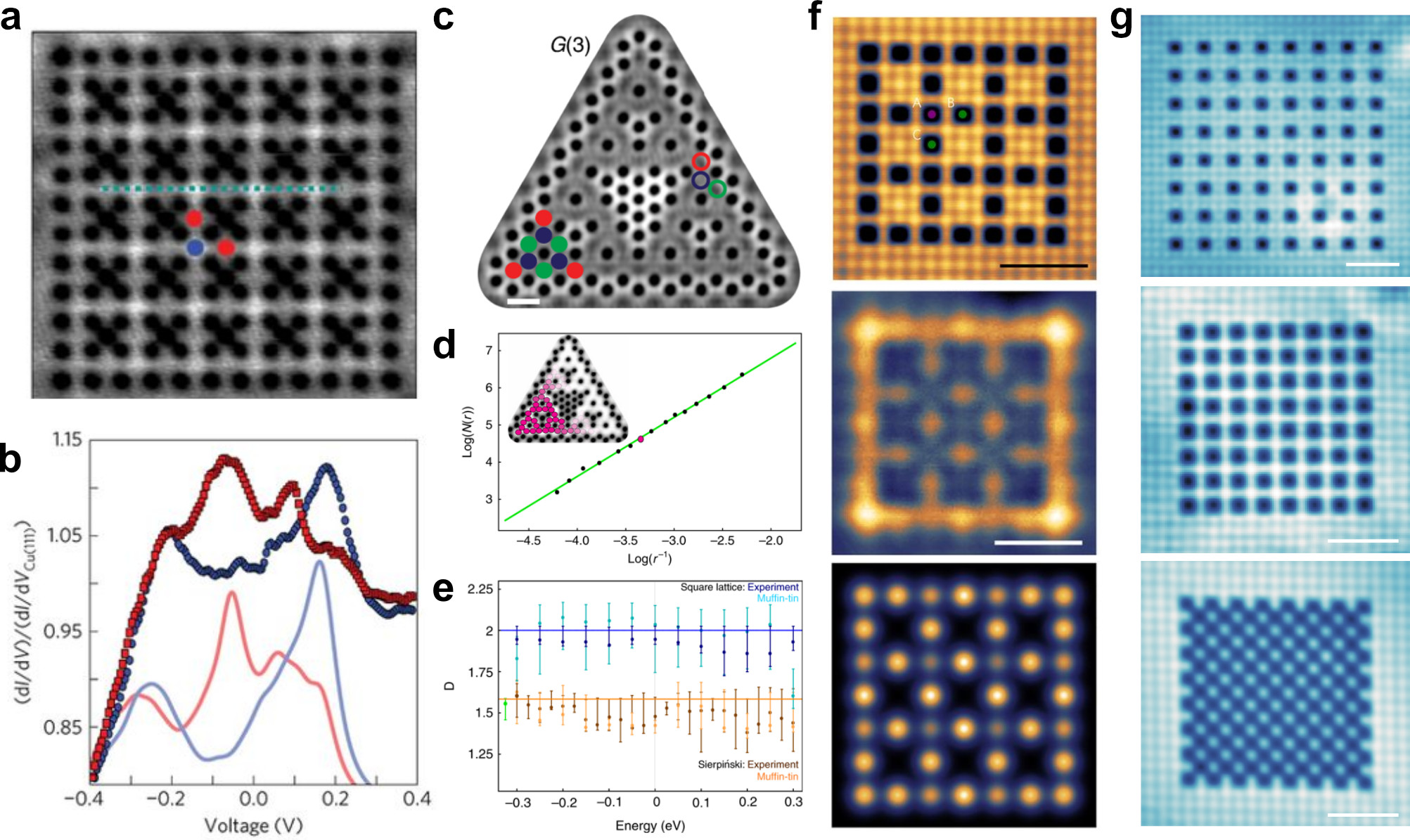}
	\caption{Experiments on artificial lattices. (a,b) Lieb lattice made with CO molecules on Cu(111) substrate (a) and the corresponding density of states measured on the different lattice sites within the unit cell (b). Adapted  by permission from Springer Nature: Ref.~\citenum{Slot2017ExperimentalLattice}, Copyright (2017). (c-e) Sierpi\'nski triangle fractal lattice constructed using CO on Cu(111) (c) and the corresponding measurement of the fractal dimension (d) and estimation of the fractal dimensions of the LDOS as a function of the energy of the G(3) Sierpi\'nski triangle (orange) and comparison with the 2D square lattice (blue) for the experimental (dark) and muffin-tin (light) wavefunction maps (e). Adapted  by permission from Springer Nature: Ref.~\citenum{Kempkes2019DesignGeometry}, Copyright (2019). (f) Lieb lattice constructed from chlorine vacancies on the c$(2\times2)$ chlorine structure on Cu(100) and the corresponding measured (middle) and calculated LDOS (bottom) maps at the energy corresponding to the flat band position. Adapted  by permission from Springer Nature: Ref.~\citenum{Drost2017}, Copyright (2017). (g) Different artificial lattice with engineered band structures constructed using the chlorine vacancy system. Adapted  by permission from SciPost: Ref.~\citenum{Girovsky2017EmergenceAtom-by-atom}, Copyright (2017).}
	\label{fig:Lieb_fractal}
\end{figure}

Using the same strategy, experiments have developeded rapidly in recent years in realizing more complicated lattice structures such as the Lieb lattice \cite{Slot2017ExperimentalLattice}, Sierpi\'nski triangle fractal \cite{Kempkes2019DesignGeometry}, Penrose tiling quasicrystal \cite{Collins2017ImagingTiling}, artificial hexagonal boron nitride \cite{YanSymmetryGraphene}, dimerized Kagome lattice \cite{Kempkes2019} and Kekul\'e lattice with edges \cite{Freeney2019}. In addition to these experimental efforts, the field is currently very active with new proposals and calculations as well \cite{Gao2014bb,Paavilainen2016bb,Li2016bb,Qiu2016bb,Allan2017bb,Ma2019bb,Qiu2019bb}. The results on the Lieb and fractal lattices are illustrated in Fig.~\ref{fig:Lieb_fractal}a,b. The band structure of a Lieb lattice consists of a Dirac cone on the corners of the first Brillouin zone with a flat band at the Dirac energy (Fig.~1h,i). As shown in Fig.~3b, the spectrum of a corner site (blue) exhibits two peaks, which can be assigned to the lowest- and highest-energy bands in the band structure of the Lieb lattice. The local density of states (LDOS) minimum between these two peaks corresponds to the Dirac point. In contrast, the spectrum of an edge site (red) exhibits a maximum in between the two band peaks, which is assigned to the flat band. Again, similar to the artificial graphene, the physics of the Lieb lattice is not restricted to the particular experimental realization and the Lieb lattice has also been constructed using chlorine vacancies by Drost et al.~\cite{Drost2017} (Fig.~\ref{fig:Lieb_fractal}f, discussed in more detail below). Finally, at higher energies, further bands stemming from the $p$-orbitals of the individual sites. The energy of these bands can be controlled in the CO/Cu(111)-system by changing the size of the lattice sites: making them larger reduces the confinement of the surface state electrons and shifts the bands down in energy. Bands arising from these higher order $p$-orbitals have been demonstrated experimentally both in Lieb and honeycomb geometries \cite{Slot2019PLattices}.

Electrons confined on a fractal lattice will behave as they move in a space having the fractal dimension $D$. This very exciting prospect has been realized in the CO/Cu(111)-system using Sierpi\'nski triangles (Fig.~\ref{fig:Lieb_fractal}c). The fractal dimension can be estimated by the box-counting dimension (Minkowski-Bouligand dimension) for an arbitrary lattice using $D=\lim _{r \rightarrow 0} \frac{\log N(r)}{\log (1 / r)}$, where $N$ is the number of circles needed to cover the LDOS and $r$ is the circles radius (Fig.~\ref{fig:Lieb_fractal}d). Plotting this dimension as a function of energy is shown in Fig.~\ref{fig:Lieb_fractal}e. Compared with the square lattice, where $D=2$ (blue solid line), the Sierpi\'nski lattice has a fractal dimension close to the theoretical Hausdorff dimension $D=1.58$ (orange solid line). Therefore, it is clear that the wavefunctions inherit the fractal dimension and the scaling properties of the confined geometry, and the dimension is non-integer (see Fig.~\ref{fig:Lieb_fractal}e). 

Instead of patterning an extended electron gas as in the case of using adsorbates on a Cu(111) surface, we could directly build the lattice out of the atomic sites hosting localized orbitals. While these approaches are largely complementary, it might be more intuitive to add some additional effects (\emph{e.g.}~magnetism) or to study edge states in lattices built from atomic building blocks. Early studies in this direction have demonstrated that 1D particle-in-a-box states can be produced in chains of adatoms  \cite{Wallis2002,Nilius2003,Folsch2004,Lagoute2005,Lagoute2007,Oncel2008}. A very intriguing subsequent development is formed by the InAs(111)A surface, which has a $2\times2$ In-vacancy reconstruction with some In adatoms adsorbed above the vacancy sites \cite{Folsch2014QuantumPrecision,Yang2011EmergentNanostructures}. The adatoms are ionized +1 donors and can confine the electrons of the InAs(111)A surface state. This forms quantum dots through electrostatic confinement. These quantum dots can be coupled and the splitting of the bonding and anti-bonding states between two separated quantum dots can be tuned smoothly by the distance. The degeneracy of the anti-bonding states can be further modified in a triple dot. Additionally, a barrier can be created reversibly by switching a surface In atom to its metastable popped-up position using the STM tip \cite{Pan2015ReconfigurableManipulation}. In this way, highly tunable electronic states could be constructed in linear chains. Another interesting platform for creating engineered electronic states out of localized orbitals was demonstrated by coupled dangling bonds  on H-Si(100) surface \cite{Schofield2013,Huff2017,Wyrick2018Atom-by-atomSilicon,Huff2018,Achal2018}. Here the physics is likely to be more complicated as attaching an electron onto the dangling bond state results in charge localization due to the formation of a polaron \cite{Huff2018}, which can be either a positive or a negative aspect depending on the targeted phenomenon.

Another appealing system is formed by Cl vacancies in a NaCl bilayer on Cu(111) \cite{Schuler2015EffectVacancies}. While a single Cl vacancy shows a localized state \cite{Repp2005ScanningJunctions}, coupled states can be formed  when two Cl vacancies are brought sufficiently close to each other. The d$I$/d$V$ maps show that the lower energy state is localized between the vacancies, while the higher energy state is stronger on the outer edge. This is a clear indication that the bonding and anti-bonding states are formed. In addition, a localized interface-state is created from the free-electron-like interface-state band of the NaCl/Cu(111). Moreover, both the vacancy state and the localized interface-state can be coupled to produce 1D quantum-well states in vacancy chains. The problem with this system, however, is that creating the vacancies requires removing single Cl-ion with the STM tip in a vertical manipulation step. Before another vacancy can be created, the Cl-ion has to be removed from the tip rendering the process time-consuming and prone to failure.

\begin{figure}[!t]
	\centering
	\includegraphics[width=.9\textwidth]{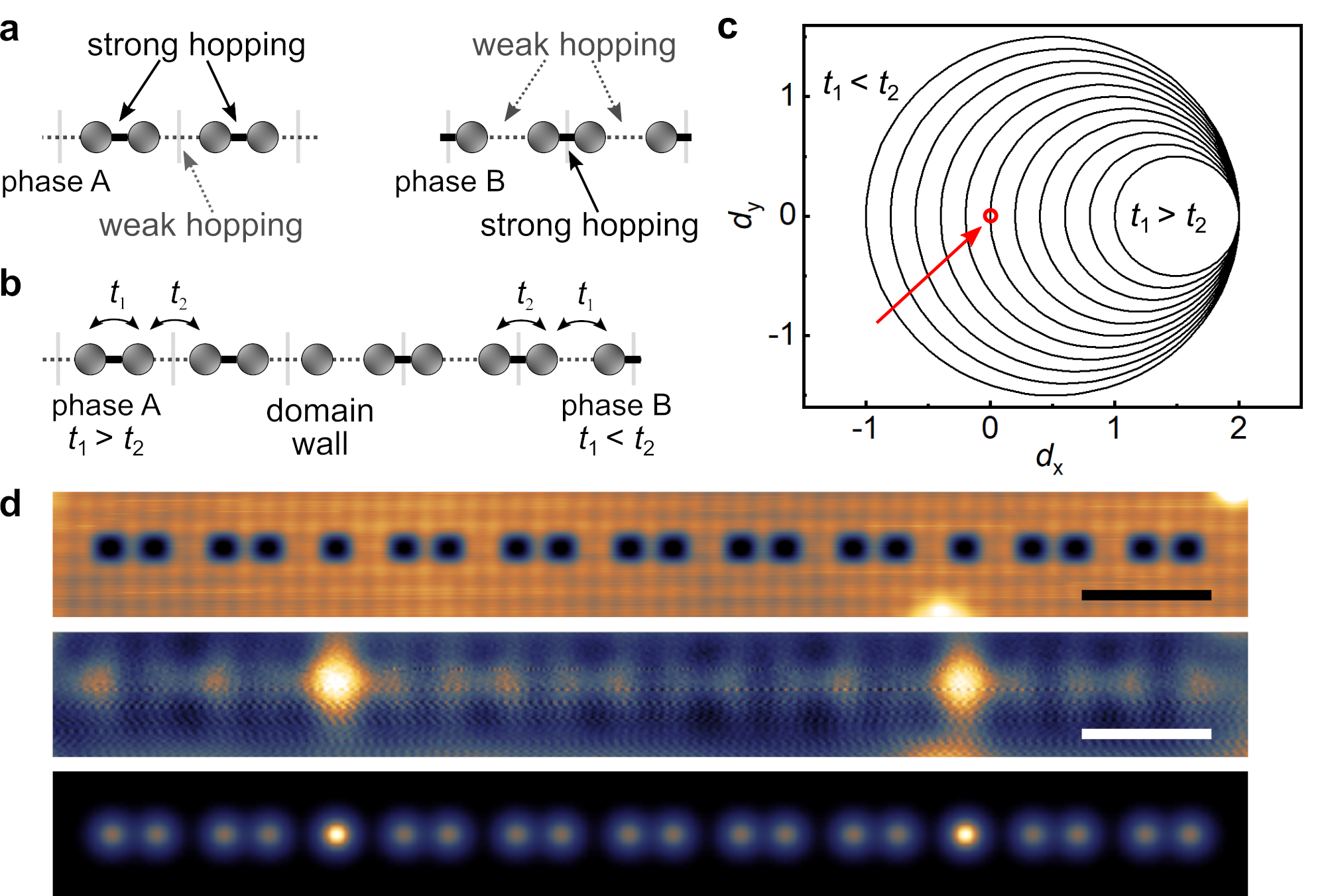}
	\caption{(a) The two phases in the SSH model where the strong hopping is either inside the unit cell (phase A) or between the unit cells (phase B). (b) Structure containing a domain wall where the intra unit cell ($t_1$) and inter unit cell $t_2$ hoppings are inverted. (c) Plots of $\mathbf{d}(k)\hat{\sigma}=d_x\hat{\sigma_x}+d_y\hat{\sigma_y}$ as a function of the $\delta t =(t_1-t_2)/2$. The red circle marks the transition between the topological and trivial phases with the associated closing of the gap. (d) Experimental realization of a structure with two domain walls using the Cl vacancy system (top) and the corresponding experimental (middle) and calculated (bottom) LDOS maps of the domain wall states. Adapted by permission from Springer Nature: Ref.~\citenum{Drost2017}, Copyright (2017).}
	\label{fig:SSH}
\end{figure}

Analogous Cl vacancy defects can be formed in the c($2\times 2$) chlorine superstructure on Cu(100) surface, with essentially error-free manipulation as illustrated by the extremely large-scale patterns  (more than 8000 bits) demonstrated in the paper by Kalff et al~\cite{Kalff2016AMemory}. Furthermore, in 2017, Drost et al.~showed that this system can be used as a platform for artificial lattices (Fig.~\ref{fig:Lieb_fractal}f~and~\ref{fig:SSH}c) \cite{Drost2017}. Cl vacancies host localized electronic states which are weakly coupled with neighbouring sites thus can be considered as an experimental representation of the tight-binding model. This model system was used to create several extended two-dimensional lattices and these works demonstrated the possibility of band structure engineering in the vacancy lattices (Fig.~\ref{fig:Lieb_fractal}f,g)  \cite{Drost2017,Girovsky2017EmergenceAtom-by-atom}. Drost et al.~realized the Lieb lattice (Fig.~\ref{fig:Lieb_fractal}f) and showed that the physics can be understood quantitatively using a simple tight-binding model with nearest- and next-nearest neighbour hoppings. Girovsky et al.~demonstrated several other lattices (some examples are shown in Fig.~\ref{fig:Lieb_fractal}g) with engineered band structures \cite{Girovsky2017EmergenceAtom-by-atom}. Using this Cl vacancy defects system on Cu(100) surface, Drost et al.~also demonstrated how to realize simple one-dimensional topological systems \cite{Drost2017} and we will first review some theoretical concepts before outlining the experimental results. 

\textbf{Topological states in one-dimensional lattices.}
While topological states can be realized by adding spin-orbit interaction to the honeycomb or kagome lattices, topological concepts can be readily illustrated in simple one-dimensional tight-binding models. Perhaps the simplest model exhibiting topological states is the one-dimensional Su-Schrieffer-Heeger (SSH) model that was developed to describe solitons in polyacetylene \cite{Heeger1988SolitonsPolymers}. The SSH model describes the physics of a dimerized one-dimensional chain within the tight-binding model and here we will explain how it can be implemented in artificial lattices and later, in graphene nanoribbons \cite{Rizzo2018TopologicalNanoribbons,Groning2018EngineeringNanoribbons}. The ideas related to dimer chains can be extended to topological domain wall states in trimers and coupled dimer chains that have also been characterized experimentally and theoretically
\cite{Cheon2015,Kim2017,Huda2018TunableChains,MartinezAlvarezEdgeLattices}.

The dimer chain (also called SSH chain) is illustrated in Fig.~\ref{fig:SSH}. The hopping parameter alternates between $t_1$ and $t_2$ for even/odd $i$ and the system has a bandgap determined by $|\delta t|=|t_1-t_2|/2$. The chain exists in two phases depending on the sign of $\delta t$ and these phases are distinguished by a topological index, the winding number. Physically, the two phases are distinguished by the location of the strong bonds. In order to understand what the ``topology'' is referring to, we can write the Hamiltonian with the help of the Pauli spin matrices
\begin{equation}\label{key}
H = \mathbf{d}(k)\hat{\sigma}=[t_1 + t_2\cos(k)]\hat{\sigma_x}+t_2\sin(k)\hat{\sigma_y}+0\hat{\sigma_z}
\end{equation}
and the winding number
\begin{equation}
\nu=\frac{1}{2 \pi i} \int_{-\pi}^{\pi} d k \frac{d}{d k} \ln (h(k))
\end{equation}
where $h(k)=d_x(k)-id_y(k)$.
If we plot the circle spanned by the vector $\mathbf{d}(k)$ as the momentum $k$ is swept over the Brillouin zone, the result depends on the sign of $\delta t=t_1-t_2$. Either the origin is contained within the circle ($\nu=1$), or it is not ($\nu=0$) (see Fig.~\ref{fig:SSH}b). What happens when we construct a system where the dimerization changes from one phase to the other, the circle naturally has to pass through the origin, \emph{i.e.} $\mathbf{d}(k)=0$. This means physically that the band gap becomes zero, which means that there will unavoidably be an in-gap state spatially localized somewhere on the domain boundary. The presence of this state is not dependent on how the domain wall is realized and it cannot be removed without doing something drastic, it is topologically protected. More on the one-dimensional topological systems can be found in several review papers and books \cite{Heeger1988SolitonsPolymers,Asboth2016AInsulators}.

The experimental realization of the SSH model using the Cl vacancies is shown in Fig.~\ref{fig:SSH}d, where the dimer chain with two domain walls is shown. The domain wall states are clearly visible in the d$I$/d$V$ map shown in the middle panel at the bias corresponding to the on-site energy. The lowest panel shows the simulated LDOS based on TB at the same energy taking into account the broadening of the levels (with a value corresponding to the experiments). This broadening gives some LDOS weight in the ``bulk'' resulting from contributions from the broadened bulk bands. The presence of the domain wall state is not dependent on its exact structure which was demonstrated by alternative realizations of a domain wall in the dimer chain \cite{Drost2017}.

\section{Graphene Nanoribbons}
\begin{figure}[!b]
	\includegraphics[width=0.72\textwidth]{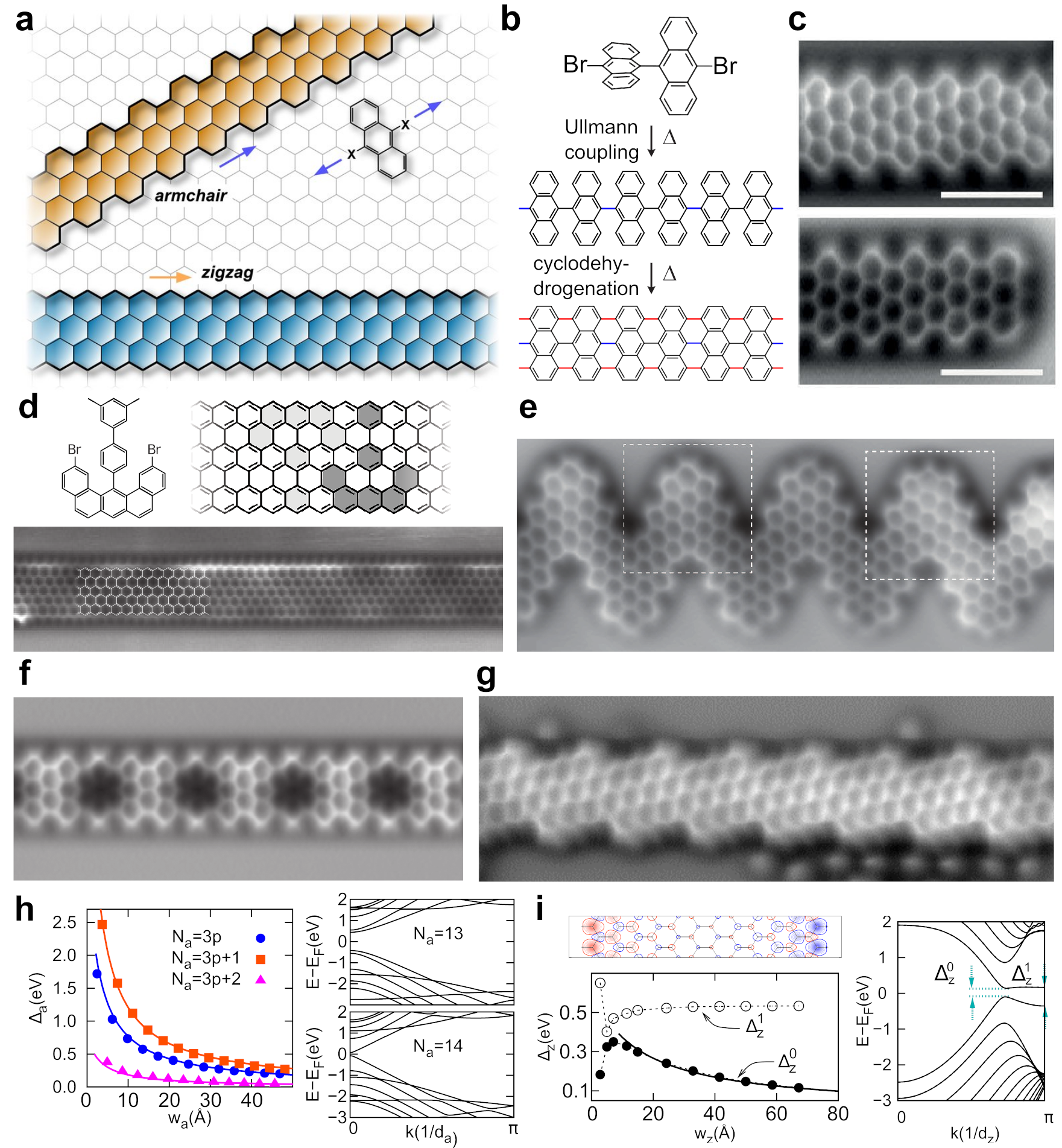}
	\caption{Nanoribbon synthesis and properties. (a) Schematic of AGNRs and ZGNRs. Adapted by permission from Springer Nature: Ref.~\citenum{Ruffieux2016On-surfaceTopology}, Copyright (2016). (b) Schematic of the Ullmann coupling route to synthesize AGNRs. (c-g) Examples of AFM images of (c) 7-AGNR (Adapted by permission from Springer Nature: Ref.~\citenum{vanderLit2013SuppressionAtom}, Copyright (2013)), (d) ZGNR (Adapted by permission from Springer Nature: Ref.~\citenum{Ruffieux2016On-surfaceTopology}, Copyright (2016)), (e) chevron GNRs (here with nitrogen edge doping \cite{Durr2018}, Adapted by permission from John Wiley and Sons: Ref.~\citenum{Marangoni2016}, Copyright (2016)), (f) boron doped 7-AGNR (Adapted  by permission from Springer Nature: Ref.~\citenum{Kawai2015AtomicallyNanoribbons}, Copyright (2015)) and (g) chiral (3,1)-GNR (Adapted with permission from Ref.~\citenum{Schulz2017PrecursorNanoribbons}. Copyright (2017) American Chemical Society). (h-i) Calculated gaps of (h) the different families of armchair GNRs (LDA level of theory) with the corresponding band structures of 13- and 14-AGNRs and (i) the ZGNRs (LDA level of theory) with the corresponding band spin density and band-structure for 12-ZGNR. Adapted with permission from Ref.~\citenum{Son2006}. Copyright (2006) by the
American Physical Society.}
	\label{fig:GNR}
\end{figure}

Graphene nanoribbons (GNRs, see schematic in Fig. 5a) are one of the best examples so far on atomically precise nanostructures. They can  be synthesized through a bottom-up chemical route on coinage metal surfaces in ultra-high vacuum (UHV). Earlier pioneering works using top-down lithography resulting in slightly disordered GNR edges have used normal lithography \cite{Han2007}, STM lithography \cite{Tapaszto2008}, nanowire etch mask \cite{Bai2009} and chemical unzipping of carbon nanotubes \cite{Kosynkin2009aa,Tao2011aa}. As even weak edge disorder is expected to affect the electronic properites of GNR \cite{Gunlycke2007,Huang2008,Stampfer2009}, using bottom-up approaches that yield atomically ``perfect'' GNRs is of crucial importance. Following the idea of on-surface polymerization by Grill et al.~\cite{Grill2007}, in 2010, Cai et al.~first successfully fabricated atomically precise armchair GNRs (AGNRs) with a width $N=7$ carbon atoms (7-AGNR) using 10,10$'$-dibromo-9,9$'$-bianthryl (DBBA) as the precursor molecules on Au(111) and Ag(111) (Fig.~\ref{fig:GNR}b) \cite{Cai2010AtomicallyNanoribbons}. The molecular precursor is first thermally evaporated onto the Au(111) or Ag(111) surface. At the annealing temperature $T_1$, polyantrylene linear chains are formed by dehalogenation and subsequent Ullmann-like  \cite{Ullmann1901UeberBiphenylreihe} coupling upon thermal activation. At a higher annealing temperature $T_2$, cyclodehydrogenation takes place and planar AGNRs are acquired. In addition to straight AGNRs (Fig.~\ref{fig:GNR}c), zigzag GNRs (ZGNRs, Fig.~\ref{fig:GNR}d)  \cite{Ruffieux2016On-surfaceTopology} and chevron-type AGNRs (Fig.~\ref{fig:GNR}e) \cite{Cai2010AtomicallyNanoribbons,Durr2018} can also be grown with the same procedure using clever design of the molecular precursors. During the past decade, various kinds of topology, doping, width and edge structure of the graphene nanoribbons have been demonstrated (examples in Fig.~\ref{fig:GNR}c-g) \cite{Talirz2016On-SurfaceNanoribbons,Fischer2017,Clair2019ControllingSynthesis}. 

Pure graphene edges are expected to undergo reconstruction \cite{Koskinen2008aa} and can be decorated with various functional groups \cite{Wassmann2008_PRL,Wassmann2010ClarsNanoribbons}. These kinds of effects are probably relevant in the case of GNRs fabricated by top-down lithography \cite{Chen2007aa} or synthesized through unzipping carbon nanotubes \cite{Kosynkin2009aa,Tao2011aa}, where the edge chemistry is poorly controlled. While the edges can be hydrogen passivated in a post-processing step \cite{Zhang2013aa}, with the GNRs obtained through the on-surface synthesis route there is no further processing necessary. The GNR edges are fully hydrogen terminated resulting from the precursor molecules. The radical sites at the ends of the GNR (the active sites for the Ullmann coupling) are eventually passivated by hydrogen present either as a contaminant in the UHV or being produced as a byproduct of the cyclodehydrogenation step.

The electronic properties of armchair and zigzag GNRs differ dramatically (Fig.~\ref{fig:GNR}h,i). The band gaps of AGNRs depend on their atomic width $N$ and can be divided into three groups (similarly to carbon nanotubes). There will be a large gap with $N = 3p+1$ ($p$ is an  integer), a medium gap with $N = 3p$, or a small (close to zero) gap with $N = 3p+2$, together with an overall decrease with $N$ within the same class (Fig.~\ref{fig:GNR}b) \cite{Nakada1996,Son2006,Son2006a,Yang2007QuasiparticleNanoribbons}. These trends can be understood by considering the discrete $k$-values in the direction perpendicular of the GNR axis and whether these values coincide with the Dirac-point of the graphene band structure. Using these arguments $N=3p+2$ family is in-fact predicted to be metallic. More realistic calculations predict edge relaxation that will open a small gap in the band structure (Fig.~\ref{fig:GNR}h). The existence of the three families and the predicted trends in the band gaps have been confirmed experimentally using the bottom-up synthesized GNRs \cite{Talirz2016On-SurfaceNanoribbons,Merino-Diez2017Width-DependentAu111}. Here, it is worth noting the ``band gap" can refer to a couple of different values depending on the system and experiment in question. In STS experiments, the resonances at positive and negative bias correspond to electron addition or removal (hole addition), respectively. This means that the STM quasiparticle gap is larger than the single-particle gap due to the polarization energies associated with electron/hole addition. It is also different from the optical gap, where one has the additional exciton binding energy. The interaction with the substrate can also result in a strong modification of the electronic states of GNRs. For AGNRs, this effect is usually weak as there is only a weak hybridization between the GNRs and the substrate \cite{Ruffieux2012ElectronicNanoribbons}, especially on Au(111). Even in this case, the screening by the underlying metallic substrate will have an effect on the STM measured band gaps and it is predicted to decrease the gap by more than 1 eV compared to the gas phase  \cite{Cai2014GrapheneHeterojunctions, Cloke2015Site-SpecificNanoribbons}. In addition, Fermi level pinning can occur when some of the GNR electronic states are situated close to the substrate Fermi level  \cite{Merino-Diez2017Width-DependentAu111,Kimouche2015Ultra-narrowNanoribbons}.
Finally, contacting a GNR to even a single metal atom by a chemical bond can strongly suppress the electron-vibron coupling without affecting the bulk electronic structure of the GNR \cite{vanderLit2013SuppressionAtom}. Moreover, strong electronic couplings have been found between the boron-doped GNR with the metal surface \cite{Pedramrazi2018ConcentrationNanoribbons,Carbonell-Sanroma2018ElectronicSurface}. Therefore, decoupling layers such as NaCl \cite{Ruffieux2016On-surfaceTopology} or intercalated Si layer \cite{Deniz2017RevealingSpectroscopy,Deniz2018ElectronicAu111} are helpful to observe the intrinsic properties of GNRs.

The ZGNRs have edge states associated with the zigzag edge that should have nearly flat dispersion and be magnetic \cite{Son2006,Son2006a}. Neglecting electron-electron interactions, tight-binding predicts degenerate flat bands. Electron-electron interactions are predicted to drive the system into a magnetic ground state and open a gap in the band structure as shown in Fig.~\ref{fig:GNR}i. The magnetic ground state has an antiferromagnetic coupling between the two edges of the ZGNR and the gap is inversely proportional to the GNR width (Fig.~\ref{fig:GNR}i). While there are indications of these edges states in non-atomically perfect sample achieved through, for example, unzipping carbon nanotubes \cite{Tao2011aa}, they have been confirmed to exist using bottom-up synthesized graphene nanoribbons \cite{Ruffieux2016On-surfaceTopology}. The zigzag edge is chemically very reactive and in fact, the intrinsic zigzag edge state cannot be resolved at all on the pure ZGNRs on Au(111). Additional functional groups that help to decouple the edges from the metallic substrate, or an ultrathin insulating film need to be used to resolve these states. Both of these strategies were used to resolve the edge state of 6-ZGNR (a ZGNR which is 6 carbon zigzag lines wide) \cite{Ruffieux2016On-surfaceTopology}. In addition to the extended zigzag edge states, the ends of the AGNRs are in the zigzag direction. They have end states associated with them and the theory suggests that they should also be spin-polarized similarly to the extended edge states \cite{vanderLit2013SuppressionAtom,Wang2016GiantEdges,Su2018a}.

The synthesis via Ullmann coupling followed by the decyclohydrogenation step works well on Au(111) and almost as well on the Ag(111) substrate. On more reactive substrates such as Cu(111) other reaction pathways are also possible. For example, the structure and formation mechanism of GNRs using DBBA molecular precursor on Cu(111) have been under debate for a while \cite{Simonov2014EffectStudy,Han2014Bottom-UpEnantioselectivity,Simonov2015CommentEnantioselectivity,Han2015ReplyEnantioselectivity,Simonov2015FromStrategy,Han2015Self-AssemblyNanoribbons}.
Recently, with the help of the unambiguous nc-AFM images, the formation of (3,1) chiral GNRs (cGNRs) (see Fig.~\ref{fig:GNR}g) on Cu(111) has been confirmed \cite{Sanchez-Sanchez2016PurelyCu111,Schulz2017PrecursorNanoribbons}. These ribbons do not form through Ullmann-coupling, the increased reactivity of the substrate enables coupling the molecules along a different direction. As the Ullmann-coupling is not operative, the same structure can be fabricated using sister non-planar molecular precursors 10,10$'$-dichloro-9,9$'$-bianthryl and halogen-free 9,9$'$-bianthryl. Substrate-independent growth of cGNRs through the Ullmann pathway can be achieved by the 2,2$'$-dibromo-9,9$'$-bianthracene molecular precursors \cite{deOteyza2016Substrate-IndependentNanoribbons,Merino-Diez2018UnravelingNanoribbons,Li2018a,Li2019a,Li2019b}.

The band alignment and the band gap can be tuned effectively through atomically-precise doping. The first examples of on-surface fabrication of doped GNRs were demonstrated in chevron-type AGNRs by one \cite{Bronner2013AligningDoping}, two \cite{Bronner2013AligningDoping,Zhang2014DirectNanoribbons} and four 
\cite{Cai2014GrapheneHeterojunctions,Vo2015Nitrogen-DopingMetamaterials} nitrogen atoms substituted per molecular precursor.  A band downshift of $0.1-0.3$ eV per dopant atom in the precursor molecule was found. Note that the nitrogen atom replaces a C--H group instead of a single carbon atom in these structures. Therefore, no significant effect on the band structure near the Fermi level is found as the nitrogen atoms carry a lone pair of electrons orthogonal to the GNR $\pi$-system, which are not donated to the carbon skeleton. On the other hand, boron-doped AGNRs have also been reported where the boron atoms are located in the centre of the AGNRs (Fig.~\ref{fig:GNR}f) \cite{Kawai2015AtomicallyNanoribbons,Cloke2015Site-SpecificNanoribbons,Pedramrazi2018ConcentrationNanoribbons,Carbonell-Sanroma2018ElectronicSurface}. Interestingly, in-gap dopant states have been shown in boron-doped AGNRs \cite{Pedramrazi2018ConcentrationNanoribbons,Carbonell-Sanroma2018ElectronicSurface}.
The fabrication of nitrile (CN) functional groups attached 7-AGNRs \cite{Carbonell-Sanroma2017DopingModification} and nitrogen, oxygen, or sulfur-doped GNRs have also been reported recently \cite{Durr2018,Zhang2017Sulfur-dopedGaps,Cao2018TuningTemperature,Rizzo2019}.

\begin{figure}[!b]
    \centering
    \includegraphics[width=0.95\textwidth]{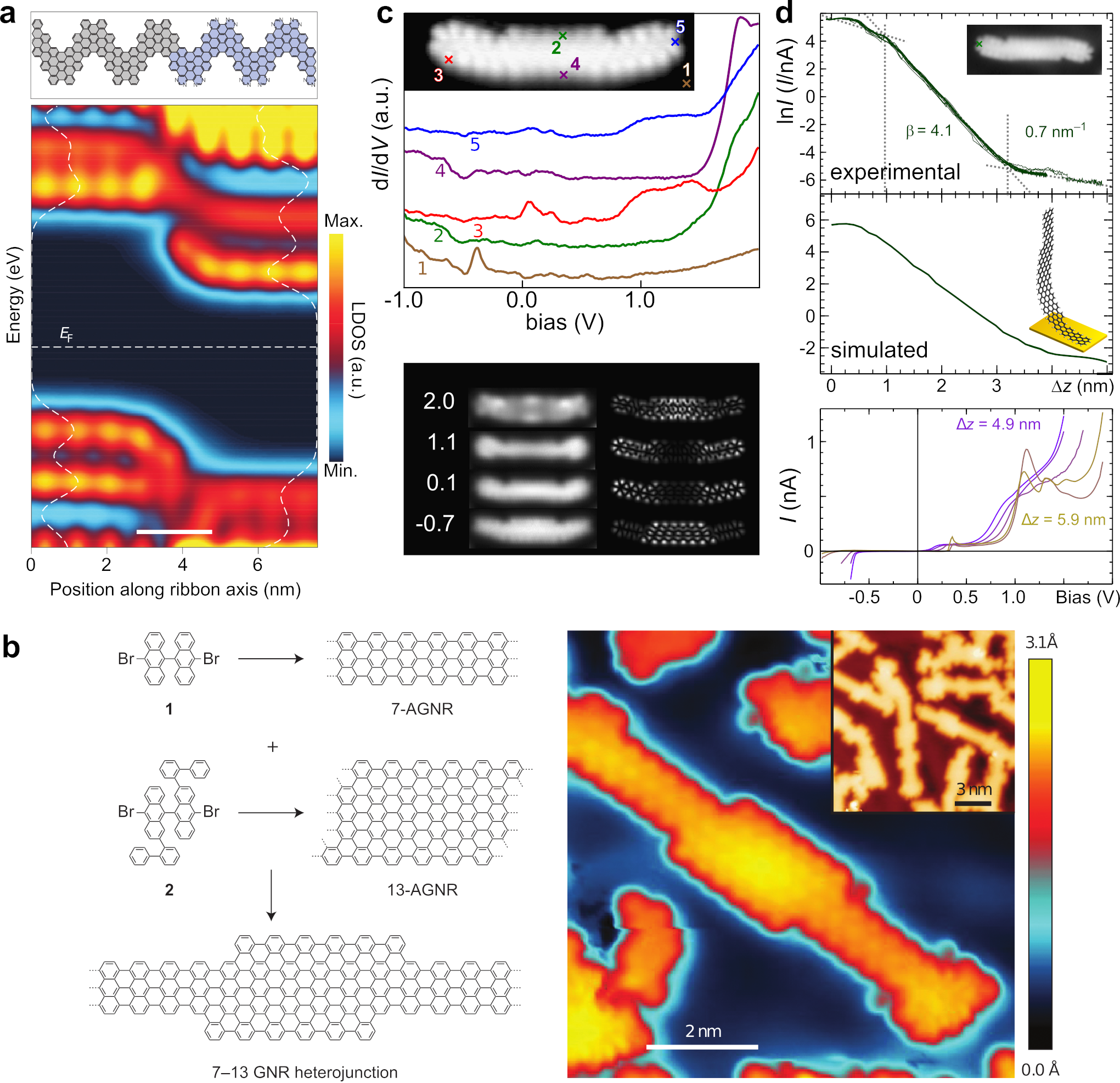}
    \caption{Experiments on GNR heterostructures.
(a) A type \uppercase\expandafter{\romannumeral2} heterojunction: pristine (left) and nitrogen-doped (right) chevron-type AGNR heterostructure and the LDOS across it. Adapted by permission from Springer Nature: Ref.~\citenum{Cai2014GrapheneHeterojunctions}, Copyright (2014). (b) A type \uppercase\expandafter{\romannumeral1} heterojunction: 7-13 AGNR heterostructure. Adapted by permission from Springer Nature: Ref.~\citenum{Chen2015MolecularHeterojunctions}, Copyright (2015).
(c,d) A metal-semiconductor junction: 5-7 AGNR heterostructure. d$I$/d$V$ spectra and LDOS maps acquired on a 5-7-5 AGNR heterostructure (c). $I-Z$ and $I-V$ curves obtained while lifting the 5-7-5 AGNR heterostructure (d). Since the ultra-narrow 5-AGNR is nearly metallic \cite{Kimouche2015Ultra-narrowNanoribbons}, the 5-7 AGNR heterostructure resembles a tunneling barrier in a metallic lead. Adapted by permission from Springer Nature: Ref.~\citenum{Jacobse2017ElectronicNanoribbon}, Copyright (2016).
}
    \label{fig:GNRhetero}
\end{figure}
\textbf{GNR heterostructures.} One direction to extend the work on atomically-precise GNRs is to increase the functionality encoded in a single GNR. This can be achieved through the formation of GNR heterostructures that can combine several different GNRs into a covalently bonded entity. Atomically  precise GNR heterostructures have been grown by on-surface synthesis (Fig.~\ref{fig:GNRhetero}). Intraribbon heterojunctions were first fabricated by temperature-induced partial cyclodehydrogenation of a surface-anchored polyanthrylene chain using only one molecular precursor \cite{Blankenburg2012IntraribbonNanoribbons}. Here, by controlling the temperature during the cyclodehydrogenation step, it was possible to control the extent of the reaction and form interfaces between fully and half cyclodehydrogenated parts of a single GNR. Another strategy using only a single precursor involves the lateral fusion of fully formed 7-AGNRs to form heterostructures between GNR segments of variable widths. Structures consisting of 7, 14, 21, and up to 56 carbon atoms across the width have been realized \cite{Ma2017SeamlessNanoribbons}.
Similar strategy has been used to form nitrogen-doped chevron-type GNR heterojunctions where the nitrogen edge-impurities bind in either five- or six-membered rings \cite{Marangoni2016,Rizzo2019}.
During the synthesis, there is also spontaneous edge reconstruction resulting in the formation of either carbazole or phenenthridine moieties. In an extended GNR, the two structures form heterojunctions.

More precisely controlled GNR heterojunctions can be constructed by combining two different precursor molecules. Chevron-type AGNR heterostructures consist of pristine (undoped) GNR and nitrogen-doped GNR behave similarly to traditional p-n junctions \cite{Cai2014GrapheneHeterojunctions}.
Figure~\ref{fig:GNRhetero}a shows the chemical structure and the related LDOS across the heterojunction of the pristine GNR (left) and nitrogen-doped GNR (right). The LDOS clearly shows the band-offset in a type \uppercase\expandafter{\romannumeral2} heterojunction. As the band offset ($\sim$~0.5~eV) occurs at an interface region of $\sim$~2~nm, the resulting electric field is $\sim$~0.2~V~nm$^{-1}$. A type \uppercase\expandafter{\romannumeral1} heterojunction has been demonstrated in a 7-13 AGNR structure (Fig.~\ref{fig:GNRhetero}b) \cite{Chen2015MolecularHeterojunctions}, since the lowest unoccupied (highest occupied) state in the 13-AGNR is lower (higher) than that in 7-AGNR. Therefore, the 7-AGNR behaves as an energy barrier of the charge carriers captured in 13-AGNR.

It is thus possible to form quantum dot states in GNR structures combining segments with different band gaps. The narrow bandgap GNR acts as the dot and the wide bandgap GNR acts as the barrier. These quantum dot states have been well studied in 7-13-7 AGNR (Fig.~\ref{fig:GNRhetero}b) \cite{Chen2015MolecularHeterojunctions}, 7-14-7 AGNR \cite{Wang2017QuantumNanoribbons} and pristine - boron-doped - pristine 7-AGNR \cite{Carbonell-Sanroma2017QuantumSubstitution} heterojunctions. Note that if the barrier length is too short, the quantum dot states are no longer localized \cite{PhysRevApplied.11.024026}. Periodic heterojunctions can be utilized in forming topological systems in analogy with the SSH model; this topic will be discussed in more detail in the next section.

Mixing nearly metallic GNR segments with wider band gap regions allows the realization of structures resembling a tunneling barrier with metallic leads (Fig.~\ref{fig:GNRhetero}c) \cite{Jacobse2017ElectronicNanoribbon}. Since  ultra-narrow 5-AGNRs are nearly metallic \cite{Kimouche2015Ultra-narrowNanoribbons},
metal-semiconductor junctions can be achieved in 5-7 AGNR heterostructures. Jacobse et al.~demonstrated by LDOS mapping that a four monomer unit long 7-AGNR segment works effectively as a tunnel barrier between nearly metallic 5-AGNR leads (Fig.~\ref{fig:GNRhetero}c) \cite{Jacobse2017ElectronicNanoribbon}. In addition to LDOS maps, it would be very desirable to be able to probe transport through these atomically well-defined GNR heterostructures. While device structures have been fabricated from single GNRs \cite{Llinas2017Short-channelNanoribbons}, this requires the sample to have monodisperse GNRs, which is not yet possible in the case of the GNR heterostructures. This limitation can be overcome as it is possible to use the STM tip to contact and lift the GNRs and subsequently, measure transport $IV$-characteristics on a suspended GNR \cite{Lafferentz2009,Koch2012}. Jacobse et al.~confirmed the picture emerging from the LDOS mapping STM experiments using this type of $IV$-measurements through suspended GNR heterostructures (Fig.~\ref{fig:GNRhetero}d), where they observed that for a sufficiently long 7-AGNR segment in between 5-AGNR ends, there was a real gap in the $IV$-characteristics. The beauty of these experiments was also in being able to carry out transport experiments on a GNR heterostructure that was first fully characterized on the atomic scale using AFM.

In the above cases, the fabrication method depends on the random process of precursor self-assembly and thus leads to a random sequence of heterojunctions. It would obviously be desirable to have more direct access to the different segments of the heterostructure.  A higher proportion of single-junction GNR heterostructures can be formed using a hierarchical GNR fabrication strategy based on the different dissociation energies of C-Br and C-I bonds \cite{Bronner2018HierarchicalHeterojunctions}. Alternatively, it is in some cases possible to complete the heterostructure formation using the STM tip \cite{Ma2017a,Nguyen2017AtomicallyPrecursor,Ma2019}. One example provided by Nguyen et al.~is through post-growth manipulation of a chevron-type AGNR resulting from a single precursor \cite{Nguyen2017AtomicallyPrecursor}. Here, voltage pulses from the STM tip result in the removal of sacrificial carbonyl groups at the tip location. This allows direct ``writing'' of the heterostructure using the STM.

\begin{figure}[!t]
    \centering
    \includegraphics[width=0.96\textwidth]{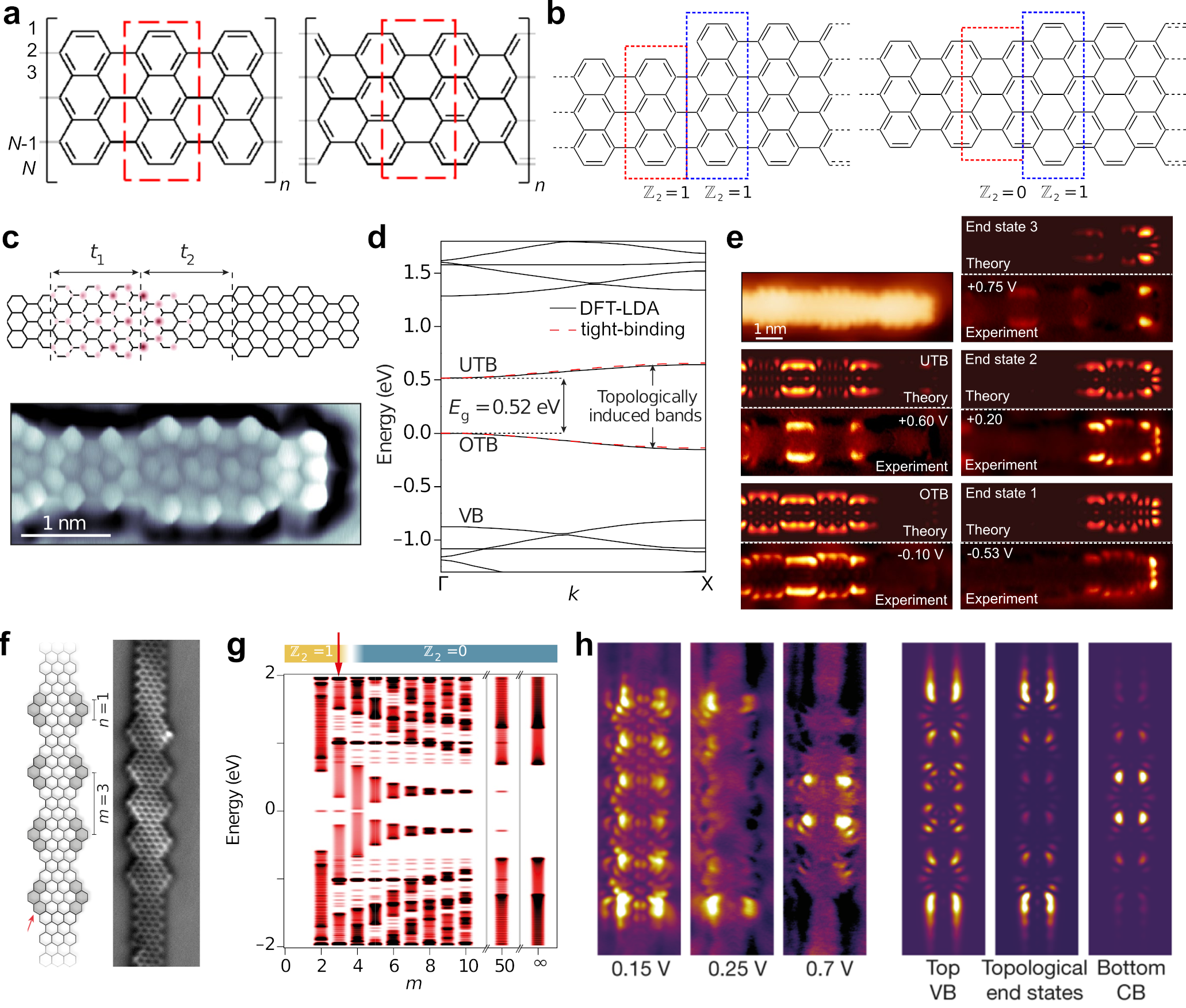}
    \caption{Theory and experiments on topological states in GNRs. (a) Different unit cells in AGNRs give rise to different topological indices.  Adapted with permission from Ref.~\citenum{Cao2017TopologicalChains}. Copyright (2017) by the American Physical Society. (b) Depending on the junction geometry, the topological index can change resulting in the formation of topological domain wall states. Adapted by permission from Springer Nature: Ref.~\citenum{Rizzo2018TopologicalNanoribbons}, Copyright (2018). (c-e) Topological states in a 7-9 AGNR superlattice. The chemical structure and high-resolution STM image (c), the calculated band structure (d) and the LDOS maps of a 7-9 AGNR superlattice (e). Adapted by permission from Springer Nature: Ref.~\citenum{Rizzo2018TopologicalNanoribbons}, Copyright (2018). (f-h) Topological states in in-line edge-extended AGNR heterostructure superlattices. The chemical structure and nc-AFM image (f), the calculated band structure (g) and the LDOS maps of an in-line edge-extended AGNR heterostructure superlattice (h). Adapted by permission from Springer Nature: Ref.~\citenum{Groning2018EngineeringNanoribbons}, Copyright (2018).}
    \label{fig:GNRtopo}
\end{figure}

\textbf{Topological states in GNRs.} Symmetry-protected topological phases in GNRs were first explored in the theory paper by Cao et al.~in AGNR \cite{Cao2017TopologicalChains} and later in cove-edged and chevron GNRs \cite{Lee2018TopologicalStates,Lin2018TopologicalSymmetries} as long as the system has a spatial (\emph{e.g.}, inversion/mirror) symmetry. In AGNRs, depending on the choice of the bulk unit cell (see Fig.~\ref{fig:GNRtopo}a, similarly to the case of the SSH chain discussed earlier), the GNR can exist in two topologically distinct phases characterized by the $\mathbb{Z}_2$ invariant \cite{Fu2007} (derived from the Zak phase \cite{Zak1989BerrysSolids,Delplace2011ZakGraphene}) being either 0 or 1. The topological states can be realized in junctions where the width changes. However, for the usual junctions (Fig.~\ref{fig:GNRtopo}b, left), the $\mathbb{Z}_2$ invariant does not change and these junctions do not host topological domain wall states. If the junction geometry is altered slightly, the unit cells change and junctions where $\mathbb{Z}_2$ changes can be realized (Fig.~\ref{fig:GNRtopo}b, right). In practise, this requires targeted design of the GNR precursor molecules and has been realized in slightly different form by two independent works \cite{Rizzo2018TopologicalNanoribbons,Groning2018EngineeringNanoribbons}.

If the junctions where the topological invariant changes are organized in a periodic array, the topological domain wall states hybridize and form bands within the original band gaps of the constituent GNRs. This is precisely the experiment carried out by Rizzo et al., who used the predicted 7-9 AGNR superlattices (Fig.~\ref{fig:GNRtopo}c,d) to generate two in-gap bands between the valence band (VB) and the conduction band (CB): the occupied topologically induced band (OTB) and the unoccupied topologically induced band (UTB) \cite{Rizzo2018TopologicalNanoribbons}. These can be understood as the bonding and anti-bonding of the topological interface states. Notably, these new bands are energetically distinct from the intrinsic band structure of the parent 7- and 9-AGNRs. At the ends of these GNRs, three additional end states emerge in this one-dimensional topologically nontrivial system: end state 1 between VB and OTB, end state 2 between OTB and UTB, and end state 3 between UTB and CB, as shown in Fig.~\ref{fig:GNRtopo}e. 

Gr\"oning et al., at the same time, showed the same physics by realizing an analogue of the Su-Schrieffer-Heeger model in staggered and in-line edge-extended AGNR heterostructure superlattices \cite{Groning2018EngineeringNanoribbons}. Fig.~\ref{fig:GNRtopo}f shows an $m=3$ in-line edge-extended AGNRs segment with 7-AGNR extensions at both ends. Here, the superlattice consists of short zigzag-segments (one indicated by the red arrow in Fig.~\ref{fig:GNRtopo}f) that are coupled. In the similar way as for the structure shown in Fig.~\ref{fig:GNRtopo}c, this creates additional bands within the band gap of the parent GNR. Tight-binding calculations show that depending on the length and spacing of these segments, the band gap between these topological bands closes and the system is driven into a topological phase. This is manifest by the domain wall states (at zero energy) between the edge-extended and regular 7-AGNR segments (Figs.~\ref{fig:GNRtopo}f,g). By synthesizing these structures, Gr\"oning et al. demonstrated experimentally the formation of the domain wall states as shown in the LDOS maps in Fig.~\ref{fig:GNRtopo}h. Further theoretical work suggests that similar structures could be used to support magnetic ordering and result in the formation of antiferromagnetic spin chains and allow the realization of Kitaev-like Hamiltonians and Majorana-type end states \cite{Klinovaja2013PRX,Rizzo2018TopologicalNanoribbons,Groning2018EngineeringNanoribbons}. These are very exciting, recent developments that strikingly demonstrate the level of control afforded by the on-surface synthesis.

\section{Outlook}
There is currently a strong push to realize designer quantum materials with electronic responses that no naturally occurring material possesses by combining different elements in atomically precise geometries. Already the current level of structural control allows precise engineering of the sample electronic structure. We used artificial lattices and graphene nanoribbons as examples of this trend, where the already existing results have demonstrated, for example, band-structure engineering and realization of topological states. It is worth noting that it is already possible to transfer the on-surface synthesized GNRs off the metal substrate onto SiO${\rm_2}$ or glass substrate \cite{Cai2010AtomicallyNanoribbons,Bennett2013,Zhao2017} followed by patterning electrical contacts to fabricate working single GNR field-effect transistors \cite{Llinas2017Short-channelNanoribbons,BorinBarin2019}. The transfer protocols still need to be improved, especially for GNRs interacting more strongly with the metal substrate (e.g.~ZGNRs \cite{Ruffieux2016On-surfaceTopology}). Eventually, embedding more functionality into the single GNRs through heterostructures \cite{Cai2014GrapheneHeterojunctions,Chen2015MolecularHeterojunctions,Jacobse2017ElectronicNanoribbon} or topological properties \cite{Groning2018EngineeringNanoribbons,Rizzo2018TopologicalNanoribbons} should allow the realization of more complex GNR devices. In the case of artificial lattices, practical applications are further into the future. However, there are already proposals on how these lattices could be used as quantum simulation platforms \cite{Khajetoorians2019_review}.

Longer term, we need to look further into the kinds of building blocks that can be brought into designer quantum materials. For example, by combining magnetism and superconductivity, it is possible to engineer one-dimensional systems that support Majorana modes, exotic particles that are their own antiparticles and that have been proposed as building blocks for a topological quantum computer \cite{Qi2011TopologicalSuperconductors,Mourik2012,Nadj-Perge2014,Sato2017}. In addition to the several works on the self-assembled atomic wires on Pb \cite{Nadj-Perge2014,Ruby2015,Feldman2016Majorana,Pawlak2016,Ruby2017}, these modes were very recently also created in an atomically precise chain formed through STM atom manipulation \cite{Kim2018}. These experiments are currently being extended to two-dimensional systems \cite{Menard2017,Palacio-Morales2018}, where it is expected that a large zoo of different topological phases can be realized \cite{Rontynen2015,Poyhonen2018}. This highlights the level of control required in reaching samples where these phenomena can be isolated, studied and engineered. This is only the beginning and we need new platforms for extending these designer concepts and discovering new physical phenomena not available in naturally occurring materials.

\section*{Acknowledgements}
This research was supported by the European Research Council (ERC-2017-AdG no.~788185 ``Artificial Designer Materials'') and Academy of Finland (Academy professor funding no.~318995 and 320555).

\bibliography{AdvPhysX-not-linked}

\end{document}